\documentclass[aps,floatfix,twocolumn]{revtex4}
\usepackage{graphics,epsfig}
\usepackage{graphicx}

\begin{document}

\title{Generating Static Fluid Spheres by Conformal Transformations}
\author{Jonathan Loranger and Kayll Lake \cite{email}}
\affiliation{Department of Physics, Queen's University, Kingston,
Ontario, Canada, K7L 3N6 }
\date{\today}

\begin{abstract}
We generate an explicit four-fold infinity of physically acceptable exact perfect fluid solutions of Einstein's equations by way of conformal transformations of physically unacceptable solutions (one way to view the use of isotropic coordinates). Special cases include the Schwarzschild interior solution and the Einstein static universe. The process we consider involves solving two equations of the Riccati type coupled by a single generating function rather than a specification of one of the two metric functions.
\end{abstract}
\maketitle

\section{Introduction}
Perhaps the simplest of all procedures that one can think of for generating new exact solutions of Einstein's equations is the use of conformal transformations. Unfortunately, when applied to vacuum, no new solutions emerge via this procedure \cite{exact}. However, when considering static fluid spheres in isotropic coordinates, the seed metric is not vacuum, but an unphysical fluid with pressure but zero energy density and viable new solutions do indeed emerge.
We now have a rather vast array of methods for generating static fluid spheres \cite{visser}, with many of the successful procedures relying on what amounts to the development of a linear equation of first order, for example \cite{es1} and \cite{es2}. In isotropic coordinates linear equations do not emerge directly \cite{lvisser}. Rather, specifying one of the two metric functions leads to a differential equation of the Riccati type. Here we do not specify either metric function but rather solve two equations of the Riccati type coupled by a single generating function. Whereas we are able to solve this system for a variety of generating functions, we have found only one class of generating functions that gives rise to tractable and physically interesting solutions of Einstein's equations.

\section{Generating Technique}
Consider spacetimes $\mathcal{O}$ that are static conformal transformations of seed metrics $\mathcal{S}$ where \cite{metric}
\begin{equation}\label{conformal}
ds_{\mathcal{O}}^2=F(r)ds_{\mathcal{S}}^2
\end{equation}
with
\begin{equation}\label{seed}
ds_{\mathcal{S}}^2=dr^2+r^2d\Omega^2-e^{2\Phi(r)}dt^2
\end{equation}
where $d\Omega^2$ is the metric of a unit sphere
($d\theta^2+sin^2(\theta)d\phi^2$) and $F$ is a freely specifiable function $>0$. We suppose that the mathematical fluid associated with $\mathcal{O}$ is generated by streamlines of constant $r,\; \theta$ and $\phi$. Since this flow is shear free, the necessary and sufficient condition for (\ref{conformal}) to
represent a static perfect fluid is given by the Walker pressure isotropy condition \cite{walker}
\begin{equation}
G_{r}^{r}=G_{\theta}^{\theta}\label{condition}
\end{equation}
where $G_{\alpha}^{\beta}$ is the Einstein tensor. The energy density is defined by $8 \pi \rho(r)=-G_{t}^{t}$ and the pressure by $8 \pi p(r)=G_{r}^{r}$ and for (\ref{condition}) we assume that $\rho+p \neq 0$ \cite{il}. Condition (\ref{condition}), along with the definitions for $\rho$ and $p$, is equivalent to the Tolman-Oppenheimer-Volkoff equation. The spacetimes $\mathcal{S}$ (for any constant $F$) do not represent physically acceptable static fluid spheres since they all have zero energy density. What we are interested in are spacetimes $\mathcal{O}$ that represent physically acceptable exact perfect fluid solutions of Einstein's equations.

\bigskip

From condition (\ref{condition}) we find
\begin{equation}\label{phipart}
r\Phi^{''}+r\Phi^{'^{2}}-\Phi^{'}+J(r)=0
\end{equation}
where
\begin{equation}\label{fpart}
J(r)\equiv(\frac{F^{'}}{F})^{'}r-(\frac{F{'}}{F})^2\frac{r}{2}-\frac{F^{'}}{F},
\end{equation}
and $\;^{'} \equiv d/dr$.
The energy density is given by
\begin{equation}\label{density}
8 \pi \rho=-\frac{F^{''}}{F^2}+\frac{3F^{'^{2}}}{4F^3}-\frac{2F^{'}}{F^2r},
\end{equation}
independent of $\Phi$, and the pressure is given by
\begin{equation}\label{pressure}
8 \pi p=\frac{2F^{'}}{F^{2}r}+\frac{3F^{'^{2}}}{4F^3}+\Phi^{'}(\frac{F^{'}}{F^2}+\frac{2}{Fr})
\end{equation}
where $\Phi$ and $F$ are linked by (\ref{phipart}). It is clear from (\ref{density}) that $F$ must have a local maximum at $r=0$ and so from (\ref{fpart}) we must have $J(0)=0$.

\bigskip

Let us specify $F$ and solve for $\Phi$ from (\ref{phipart}). The formal solution is given by
\begin{equation}\label{methoda}
\Phi=\int b(r) dr +C
\end{equation}
where
\begin{equation}\label{j}
b^{'}+b^2-\frac{b}{r}+\frac{J}{r}=0
\end{equation}
and $C$ is a constant.
Since equation (\ref{j}) may be solved analytically only for certain $J$, we can ask what $F$ gives rise to this particular $J$? The answer follows from (\ref{fpart}) and is given by
\begin{equation}\label{f}
F=exp\left(\int \tilde{b}(r)dr+\tilde{C}\right)
\end{equation}
where
\begin{equation}\label{btilda}
\tilde{b}^{'}-\frac{\tilde{b}^2}{2}-\frac{\tilde{b}}{r}-\frac{J}{r}=0
\end{equation}
and $\tilde{C}$ is a constant. Alternatively, we can specify $\Phi$ and solve for $F$. This is equivalent to (\ref{f}) with (\ref{btilda}) where we consider $J$ generated from (\ref{phipart}). Again, we can ask what $\Phi$ generated a particular $J$. The solution is given by (\ref{methoda}) with (\ref{j}).
In either case our ability to proceed revolves around our ability to solve Riccati equations of the type (\ref{j}) and (\ref{btilda}).

\bigskip

In the usual way, equations (\ref{j}) and (\ref{btilda}) can be written in linear form as
\begin{equation}\label{psi}
    \psi^{''}-\frac{\psi^{'}}{r}+\frac{J \psi}{r}=0
\end{equation}
and
\begin{equation}\label{psit}
    \tilde{\psi}^{''}-\frac{\tilde{\psi}^{'}}{r}+\frac{J \tilde{\psi}}{2 r}=0
\end{equation}
where, up to a scale factor in $t$ and a constant conformal factor, $\psi=e^{\Phi}$ and $\tilde{\psi}=1/\sqrt{F}$. Rather than specify $F$ or $\Phi$, here equations (\ref{psi}) and (\ref{psit}) are solved simultaneously, coupled by the generating function $J$. The solutions for $\psi$ and $\tilde{\psi}$ are, of course, quite different in general and, as explained below, distinct even for $J=0$.
\section{Solutions to the Riccati System}
Given a particular solution to a Riccati type equation, standard procedures \cite{poly} allow the construction of more general solutions. However, this procedure for generating solutions usually starts from very simple particular solutions, and we have found no non-trivial known solutions applicable to the system (\ref{psi}) and (\ref{psit}). Rather, what we have done is to use the computer algebra system Maple \cite{maple} to generate solutions to this system.

\bigskip

To motivate our choice for $J$, first consider
\begin{equation}\label{firstf}
F=\frac{A}{(1+Br^2)^n}
\end{equation}
where $A$ and $B$ are constants and $n$ is a ratio of integers. This gives
\begin{equation}\label{specialj}
J=\frac{2(2-n)nB^2r^3}{(1+Br^2)^2}
\end{equation}
which distinguishes two special cases for which $J=0$: $n=0$ and $n=2$. It is important to note that (\ref{firstf}) is but a special case that leads to (\ref{specialj}). With $J=0$, $e^{\Phi}=C+Dr^2$ where $C$ and $D$ are constants. The cases $n=0$ are physically unacceptable, since, as explained above, the associated energy density vanishes. All cases with $n=2$ are conformally flat and so represent the well-known Schwarzschild interior solution \cite{buch}. A special case is given by $C=1$ and $D=B$ which is the Einstein static universe. (A cosmological constant $\Lambda=4B/A$ can be introduced to give zero pressure.)

\bigskip

Motivated by the foregoing, we have considered the generating functions
\begin{equation}\label{generalj}
J=\frac{2(2-n)nB^2r^b}{(1+Br^2)^a}
\end{equation}
where $a$ and $b$ are integers, and have been able to solve the Riccati system (\ref{psi}) and (\ref{psit}) analytically for the integers shown in Table \ref{Solutions}.

\begin{table}[ht]
\caption{\label{Solutions}Analytic Solutions}

\begin{tabular}{|c|c|}

  {a} & {b}\\
  \hline
  1 & 1, 3, 5\\
  2 & 1, 3, 5, 7\\
  3 & 1, 3, 5\\
  4 & 1, 3, 5,7\\
  5 & 3, 5\\
  6 & 3, 5, 7\\

\end{tabular}
\end{table}

\bigskip

However, solving the differential equations does not mean that we can find a physically acceptable, or even tractable, solution to the Einstein equations. In some cases (e.g. $a=2, b=5$) we have been unable to construct the associated energy density and pressure simply due to the complexity of the background spacetime. In other cases (e.g. $a=1, b=3$) the most elementary physical requirements cannot be met (finite positive $\rho$ and $p$ at the origin $r=0$ with monotone decreasing values outward). Of the solutions represented in Table \ref{Solutions} we have found only one case of physical interest: ($a=2, b=3$), that is, (\ref{specialj}).

\section{Physically Acceptable Solutions}
Now starting with (\ref{specialj}) from (\ref{psi}) and (\ref{psit}) we find
\begin{equation}\label{generalphi}
e^{\Phi}=\mathcal{C}_{1}(1+Br^2)^{(1+\sqrt{N})/2}+\mathcal{C}_{2}(1+Br^2)^{(1-\sqrt{N})/2}
\end{equation}
where
\begin{equation}\label{N}
N \equiv 2n^2-4n+1,
\end{equation}
and
\begin{equation}\label{generalf}
F=\frac{1}{(\mathcal{C}_{3}(1+Br^2)^{n/2}+\mathcal{C}_{4}(1+Br^2)^{1-n/2})^2}
\end{equation}
where the $\mathcal{C}_{x}$ are constants. From (\ref{N}) we have $n \geq 1+\sqrt{2}/2$ and $n \leq 1-\sqrt{2}/2$. Whereas the metric is remarkably simple, the resultant expressions for the energy density and isotropic pressure are very long and not reproduced here. We resort to graphical demonstrations.

\bigskip

\begin{figure}[ht]
\epsfig{file=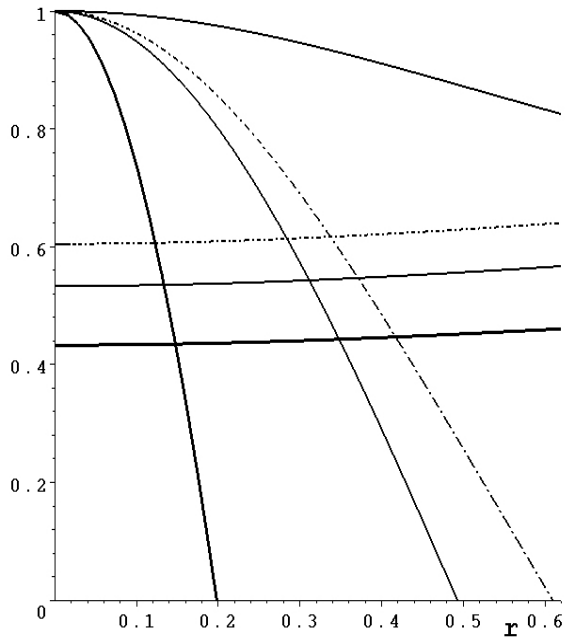,height=3.5in,width=3in,angle=0}
\caption{\label{c1c2var}Variation of $\mathcal{C}_{1}$ and $\mathcal{C}_{2}$. For all the curves $\mathcal{C}_{3}=1$, $\mathcal{C}_{4}=4$, $n=7/4$ and $B=1$. The top curve shows $\rho/\rho(0)$. The curves which intersect the abscissa give $p/p(0)$ and the horizontal curves show the square of the adiabatic sound speed. The configurations terminate and match onto vacuum at $p=0$. Two sets of curves, which coincide, are shown. For one $\mathcal{C}_{1}=1$ and $\mathcal{C}_{2}$ is varied: 1/2 for the thick curves, 1/4 for the regular curves and 1/8 for the dashed curves. For the other,  $\mathcal{C}_{2}=1/4$ and $\mathcal{C}_{1}$ is varied: 1/2 for the thick curves, 1 for the regular curves and 2 for the dashed curves. }
\end{figure}

The fact that the energy density is unaffected by $\Phi$ is demonstrated in Figure \ref{c1c2var} where we have varied $\mathcal{C}_{1}$ and $\mathcal{C}_{2}$. Two sets of curves shown coincide. This degeneracy arises due to the fact that we have not set a scale for $t$.

\begin{figure}[ht]
\epsfig{file=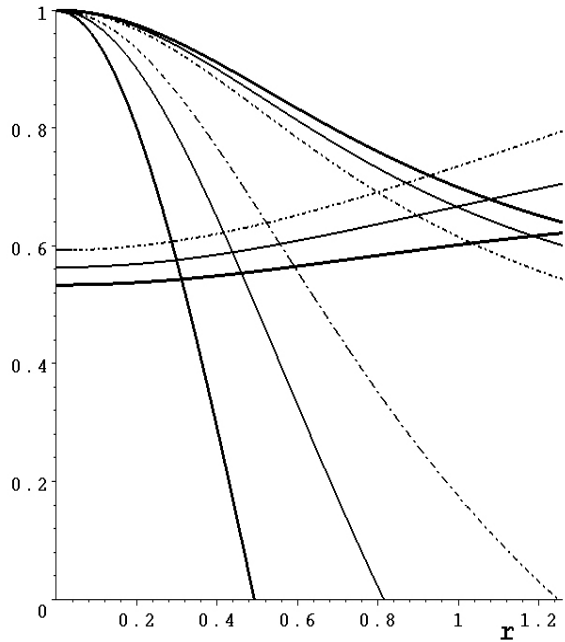,height=3.5in,width=3in,angle=0}
\caption{\label{c3var}Variation of $\mathcal{C}_{3}$ and $\mathcal{C}_{4}$. For all the curves $\mathcal{C}_{1}=1$, $\mathcal{C}_{2}=1/4$, $n=7/4$ and $B=1$. The nature of the curves can be recognized from Figure \ref{c1c2var}. For one set of curves $\mathcal{C}_{4}=4$ and $\mathcal{C}_{3}$ is varied: 1 for the thick curves, 3/4 for the regular curves and 1/2 for the dashed curves. For the other set $\mathcal{C}_{3}=1$ and $\mathcal{C}_{4}$ is varied: 4 for the thick curves, $\sim$5.3 for the regular curves and 8 for the dashed curves.}
\end{figure}

In Figure \ref{c3var} we have varied $\mathcal{C}_{3}$ and $\mathcal{C}_{4}$. Again, two sets of curves shown coincide. This degeneracy arises due to the fact that the essential physics does not change under a constant conformal transformation.

\section{Discussion}

An explicit four-fold infinity of new physically acceptable exact perfect fluid solutions of Einstein's equations have been generated by solving two equations of the Riccati type coupled by a single generating function rather than specifying one of the metric functions. Special cases of these solutions include the Schwarzschild interior solution and the Einstein static universe. The solutions are qualitatively similar to the Tolman IV solution (see for example \cite{es2}) and so should be of interest for the study of internal properties of neutron stars \cite{lat}.

\bigskip

\begin{acknowledgments}
KL is supported by a grant from the Natural Sciences and
Engineering Research Council of Canada. Portions of this work were made possible by use of \emph{GRTensorII} \cite{grt}.
\end{acknowledgments}


\begin{thebibliography}{}
\bibitem{metric}We use geometrical units and usually designate functional dependence only on the first appearance of a function.
\end{thebibliography}
\end{document}